# Time-domain characterization of electric field vector in multi-terahertz pulses using polarization-modulated electro-optic sampling

**NATSUKI KANDA**[1,2,*], **MAYURI NAKAGAWA**[1], **YUTA MUROTANI**[1], **AND RYUSUKE MATSUNAGA**[1]

[1]*The Institute for Solid State Physics, The University of Tokyo, Kashiwa, Chiba 277-8581, Japan.*
[2]*Research Center for Advanced Photonics, RIKEN, 2-1 Hirosawa, Wako, Saitama, 351-0198, Japan.*
*\* n-kanda@riken.jp*

**Abstract:** We demonstrated characterizing the electric field waveform of multi-terahertz pulses (10–50 THz) as vector quantities in the time domain by applying the polarization modulated electro-optic sampling (POMEOS) method. The problem of an ultrabroadband gate pulse was solved by modifying the fitting function in POMEOS and its validity was confirmed through numerical simulations. High accuracy and precision of approximately 1 mrad with 3 s accumulation were demonstrated. Our method can be applied not only to multi-terahertz polarization measurements for linear response but also to the evaluation of the driving field of intense pulses for nonlinear response or material control.

## 1. Introduction

Recently, laser-based coherent pulsed light sources with a fixed carrier-envelope phase have been developed in the 10–50 terahertz (THz) range, which lies between the THz and mid-infrared regions and is often called the multi-THz region [1−7]. The detection of the time-domain waveforms of the electric field is possible using ultrashort optical pulses with a pulse width of approximately 10 fs, in a similar configuration to the conventional THz time-domain spectroscopy (TDS) around the 1 THz region. Such a multi-THz TDS is highly beneficial for measuring the broadband and complex response functions of materials and the ultrafast dynamics in the pump-probe scheme [8,9]. More importantly, the response function of a material is generally a tensor quantity. Multi-THz polarization-resolved measurements are required to obtain the off-diagonal components that reveal vibrational circular dichroism or anomalous Hall conductivity. Moreover, recently the ultrafast control of material properties by utilizing the time-varying polarization degrees of freedom in intense multi-THz light pulses has attracted tremendous attention for controlling topological materials [10−16], nonlinear phononics [17,18], and chiral phonons [19,20]. Therefore, there is a high demand for an evaluation technique for the vector field waveforms of multi-THz pulses.

Conventionally, polarization-sensitive measurements have been implemented using polarization-modulated Fourier transform infrared (FTIR) spectroscopy by using a photoelastic modulator (PEM) for the mid-infrared region [21,22]. Despite the advantages of high frequency resolution and a broad spectral range, it is difficult to apply FTIR to time-resolved measurements on an ultrafast femtosecond time scale. It is also impossible to determine the time evolution of the electric field vector because the phase is not detected.

Another challenge for polarization-resolved spectroscopy in the multi-THz region is the absence of necessary optical elements. For approximately 1 THz, polarization-resolved TDS can be performed with free-standing wire-grid polarizers (WGPs) by simple projection in two directions [23−25] or by modulating the signal by the high-speed rotation of a polarizer to eliminate artifacts [26]. However, it is difficult to fabricate free-standing WGPs in the multi-THz region because of much shorter wavelengths. Although WGPs are available on substrates such as Thallium Bromo-Iodide (KRS-5), the group velocity dispersion of the substrate severely distorts the waveform. Apart from WGPs, there are other approaches, such as multicontact photoconductive antennas [27,28] and air-biased coherent detection [29,30]. However, these methods also have limitations in terms of the bandwidth or intensity of the laser, as well as difficulty with cross-talk effects.

In this work, we applied the polarization-modulated electro-optic sampling (POMEOS) method, which was developed in THz-TDS around 1 THz [31–33], to measure the waveforms of multi-THz pulses as vector values. In POMEOS, modulating the polarization of the optical gate pulse generates the modulated electro-optic (EO) signal, which allows us to directly determine both the amplitude and azimuthal angle of the electric field without any polarizers. However, the extension of POMEOS to the multi-THz region was not obvious because the 10-fs-class ultrashort pulse contains a broadband spectrum, where the retardation by the polarization modulator has a wavelength dependence. To solve this problem, we modified the fitting function of the modulated signals. Owing to this improvement, the time-domain characterization of the multi-THz electric field vector at 10–50 THz was successfully demonstrated by modulating the broadband gate pulse. Based on the numerical simulation, we delve into the adjustments made to POMEOS for the 1-THz region. We also address the prerequisites for accurate measurements in the multi-THz range. These results open a new avenue for ultrafast science in the multi-THz region.

## 2. Experiments and Results

A Yb:KGW regenerative amplifier (PH2-2mJ-SP, Light Conversion) with a center wavelength of 1030 nm and a pulse width of 160 fs was used as the light source. A part of the output was compressed to less than 13 fs with a pulse energy of 60 µJ, by using multi-plate broadening and dispersion compensation [7,34,35]. The beam is split into two beams with a controlled time delay for multi-THz TDS [7,8]. A 10-µm-thick GaSe crystal with ab-plane was used for generating broadband multi-THz pulses which cover 10–50 THz by intra-pulse difference frequency generation process. Another 10-µm-thick GaSe crystal was used for the detection of the electric field of multi-THz pulses with the EO sampling (EOS) method.

In conventional EOS, linearly polarized gate pulses with fixed polarization directions enter the EO crystals. Through propagation in the detection crystal, the gate pulse acquires ellipticity proportional to the multi-THz electric field, which is measured by balanced detection using a quarter-wave plate (QWP) and a polarization beam splitter. In contrast, in this study, the azimuthal angle of the linearly polarized gate pulses before the EO crystal was modulated by a PEM (I/FS50, HINDS Instruments) and achromatic QWP, as shown in Fig. 1(a). The PEM induces birefringence in the crystal, which modulates the ellipticity of the gate pulse. The ellipticity was subsequently converted to a modulation of the azimuthal angle of the linear polarization using a QWP whose optical axis was set parallel to the initial polarization. The phase retardation of the PEM can be expressed as $\delta = \delta_0 \sin(2\pi f_{\text{PEM}} t)$, where $\delta_0$ is the modulation depth, $f_{\text{PEM}}$ is the modulation frequency, and $t$ is the laboratory time. After the QWP, the polarization azimuthal angle $\psi$ becomes half of $\delta$. When the EO crystal has three-fold rotational symmetry, the modulated EO signal can be expressed as $E_{\text{THz}} \sin(2\psi + \theta)$, where $E_{\text{THz}}$ and $\theta$ are the amplitude and the azimuthal angle of the multi-THz electric field vector, respectively. In total, the EO signal as a function of the laboratory time $S(t)$ becomes

$$S(t) = E_{\text{THz}} \sin(\delta_0 \sin(2\pi f_{\text{PEM}} t) + \theta). \quad (1)$$

Figure 1(b) shows a schematic of the retardation of the PEM and polarization state of the gate pulses in the laboratory time. If the repetition rate of the laser $f_{\text{rep}}$ is much higher than $f_{\text{PEM}}$, the modulated EO signal can be easily fitted using Eq. (1). However, $f_{\text{rep}}$ of a regenerative amplifier is often lower than $f_{\text{PEM}}$, e.g., $f_{\text{rep}} = 3$ kHz and $f_{\text{PEM}} \sim 50$ kHz in our case. In such a case, the EO signal is not a smooth function of the laboratory time because of aliasing, as shown in Fig. 1(c). In a previous study [31], $f_{PEM}$ was locked to $m/n \times f_{rep}$ to resolve the aliasing, where $m$ and $n$ are small integers. By contrast, we measured the EO signal $S$ and the PEM reference signal (the square wave of $f_{PEM}$) for each laser pulse simultaneously for free-running $f_{rep}$ and $f_{PEM}$. By post data analysis, the PEM phase, which corresponds to $2\pi f_{PEM} t$, was determined by the phase of reference signal at trigger timing. The PEM phases fill the whole range of 0–$2\pi$ with several tens of laser pulses because of a moderate aliasing, as shown in Fig. 1(d). The EO signal can then be plotted as a smooth function of the PEM phase, as shown in Fig. 1(e). An optical chopper with a modulation frequency of $f_{rep}/2$ was inserted to eliminate artifacts caused by residual birefringence in the EO crystal. The red and blue points in Figs. 1(c)–(e) correspond to the data with and without the multi-THz field, respectively. The latter shows the effect of the residual birefringence, which must be removed before the fitting analysis. To determine the difference between the data with and without the multi-THz field, a dataset with the same PEM phase was necessary. For this purpose, the PEM phases were split into 60 segments and the pulse-resolved data (dots in Fig. 1(e)) were averaged for each segment. Subsequently, these representative points were interpolated into evenly spaced PEM phases, as indicated by the circles in Fig. 1(e). After averaging and interpolation, the difference signal caused by the multi-THz field was obtained, as shown in Fig. 1(f). The data were well fitted by the function $a_0 + E_{\text{THz}} \sin(\delta_0 \sin(\varphi - \varphi_0) + \theta)$, which is a modified function of Eq. (1). The offset term $a_0$ originates from the wavelength dependence of the modulation depth of the PEM, because the gate pulse has a large bandwidth from 900 to 1100 nm [7]. From the fitting, $E_{\text{THz}}$ and $\theta$ were determined at each delay time between the multi-THz and gate pulses. The fitting function includes five parameters: $a_0, E_{\text{THz}}, \delta_0, \varphi_0,$ and $\theta$. Among these, $\delta_0$ and $\varphi_0$ depend only on the PEM setting. These were fixed to the values obtained by fitting certain data, typically at the peak amplitude. Therefore, there remained only three fitting parameters, $a_0, E_{\text{THz}},$ and $\theta$, when the delay time of the gate pulse was scanned. From $E_{\text{THz}}$ and $\theta$ at each delay time, the waveform of multi-THz pulses was obtained as vector values. The angle of $\theta = 0$ was defined by measuring a field with a known polarization, typically horizontal polarization. Figure 1(g) shows the experimental results for the waveform of a broadband multi-THz electric pulse generated by GaSe. Horizontally polarized monocycle pulses were successfully measured. Figure 1(h) shows the power spectra of $x$ and $y$ components for the same pulse, demonstrating broadband measurements covering 10–50 THz.

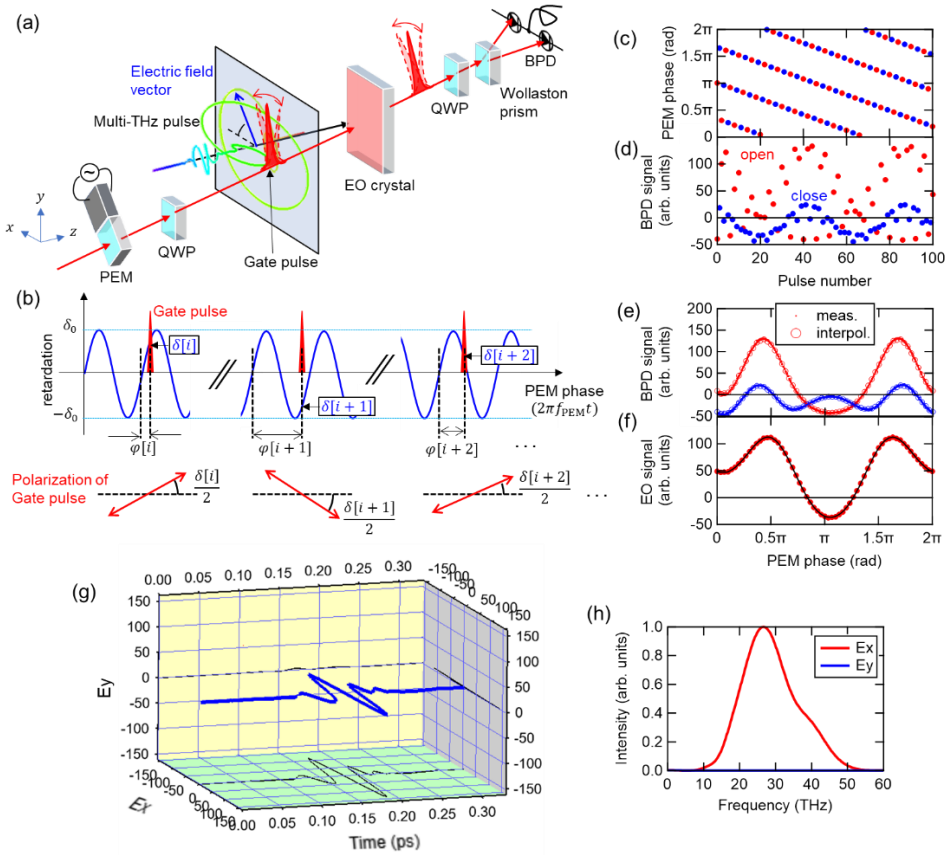

Fig. 1. (a) Schematic of POMEOS setup. (b) PEM phase and polarization of gate pulse for each pulse. Because of the relationship between modulation frequency and repetition rate, aliasing occurs. (c, d): An example of raw data of BPD signal (c) and PEM phase (d) for each laser pulse. Blue: chopper closed, Red: chopper opened. (e) Corresponding data as functions of PEM phase. Dot: measured data, circle: interpolated data. (f) Subtracted signal where the effect of residual birefringence is eliminated. Red circle: measured, black line: fitted. (g) Experimental result for a monocycle, broadband multi-THz pulse. (h) The power spectra for the waveform in (g). Red and blue show $x$ and $y$ components.

To confirm the characterization of the arbitrarily polarized pulses, we measured other polarization directions and ellipticities. First, linear polarization was tested: The GaSe crystal for multi-THz generation was rotated by 10°/step, so as to rotate the polarization plane by 30°/step. To ensure high polarization degree, a WGP on a KRS-5 substrate was inserted and rotated by 30°/step simultaneously. The results are presented in Figs. 2(a)–(f). In all polarization directions, linear polarization was effectively measured using our method. Second, a QWP for 33.3 THz was inserted to generate a circular polarization. Figure 2(g) shows the results when the angle between the incident polarization and the fast axis is 45°. The electric field was rotated clockwise along the $xy$-plane. We also observed a counterclockwise rotation when the QWP was rotated to –45° (Fig. 2(h)). Figure 2(i) shows the power spectra of the right- and left-circular polarization components of the pulse in Fig. 2(g).

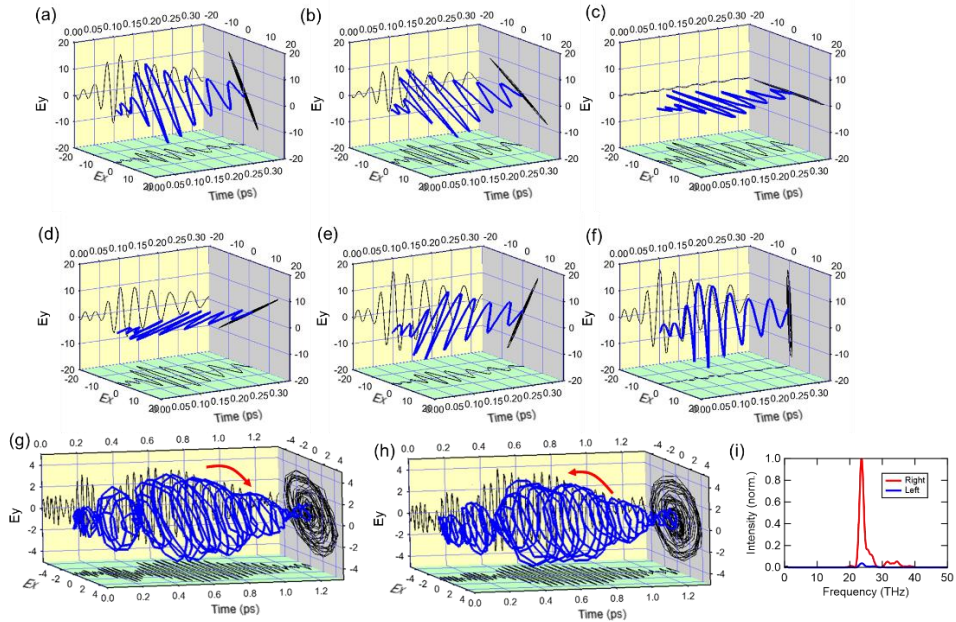

Fig. 2. Results of the waveforms of electric field vectors. (a–f) Linearly polarized multi-THz pulses with 30° rotation step. (g, h) Circularly polarized pulse. Red arrows show the direction of the rotation of the field. (i) Power spectra of circular polarization in (g). Red and blue show right- and left-circular components, respectively.

A right-handed component that was much larger than the left was successfully identified. The spectral peak at approximately 23 THz originates from a combination of the incident spectrum and the transmittance of the QWP.

Next, the performance of polarization measurements was evaluated using linearly polarized multi-THz pulses. To evaluate precision, we obtained the statistics of 100-times-repeated measurements at a fixed time delay. Figure 3(a) shows a histogram of the measured direction of the electric field. The accumulation time for each measurement was 1 s, which contained 3,000 laser pulses. The data followed a normal distribution, with a standard deviation of 1.7 mrad. The dependence on the accumulation pulse number is shown in Fig. 3(b). The standard deviation decreased in proportion to the inverse square root of the accumulation number, which is consistent with a normal distribution. The precision reached 1 mrad for 9,000 laser pulses (3 s). The accuracy was also evaluated by comparing the angle of the WGP and the detected field direction, as shown in Figs. 3(c)–(e), for different angle ranges. From Fig. 3(c), the angle was accurately measured in a wide range from –90° to +90°, as also confirmed in Figs. 2(a)–(f). As shown in Figs. 3(d) and (e) with enlarged scales, the measurement was also accurate in the small-angle ranges on the order of 1 mrad.

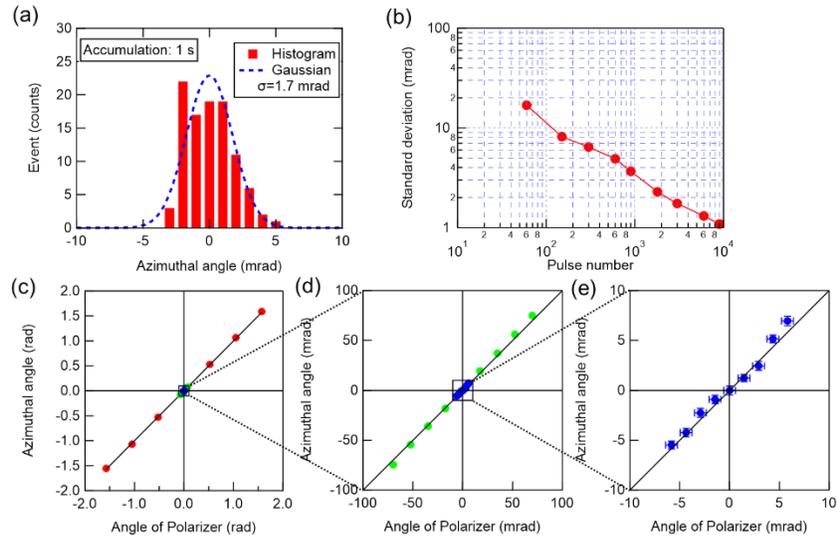

Fig. 3. Precision and accuracy. (a) Histogram of the measured azimuthal angles. The distribution is well fitted by Gaussian function (blue dashed line). (b) Accumulation pulse number dependence of azimuthal angle precision. (c–e) The relationship between measured azimuthal angles and known WGP angles in different angle ranges.

## 3. Discussion

By applying POMEOS to an ultrashort gate pulse with a duration of approximately 10 fs, we successfully demonstrated the measurements of multi-THz time-domain waveforms as vector values. However, the fitting function required an additional offset term $a_0$ compared to a previous THz study [31]. Figure 4(a) shows an example of the obtained fitting parameters, which depend on the delay time. As indicated by the green curve, a finite $a_0$ is necessary for better fitting. To explain this, we must go beyond the ideal situation, in which only the azimuthal angle is modulated without ellipticity. This is only the case when the retardation of the QWP and PEM is uniform over the entire gate pulse bandwidth. However, their wavelength dependence may affect the measurements in actual cases because our gate pulse has a broad bandwidth.

To evaluate this hypothesis, numerical simulations were conducted. Instead of the Pockels effect formalism often adopted for the EOS in the 1-THz region, we describe the interaction between the gate and multi-THz pulses as sum and difference frequency generation (SFG and DFG) processes so that the finite frequencies of the multi-THz pulses are taken into account. The second-order nonlinear susceptibility in a three-fold symmetric system has only 4 non-zero components with one independent parameter; $\chi_{yyy} = -\chi_{yxx} = -\chi_{xxy} = -\chi_{xyx}$. The transmitted gate pulse after the EO crystal $\boldsymbol{E}^{g,\text{out}}(\omega)$ is expressed as

$$\begin{bmatrix} E_x^{g,\text{out}}(\omega,\tau) \\ E_y^{g,\text{out}}(\omega,\tau) \end{bmatrix} = \begin{bmatrix} E_x^g(\omega) \\ E_y^g(\omega) \end{bmatrix} + i\omega b \int d\Omega \left( e^{i\Omega\tau} \begin{bmatrix} -E_y(\Omega) & -E_x(\Omega) \\ -E_x(\Omega) & E_y(\Omega) \end{bmatrix} \begin{bmatrix} E_x^g(\omega-\Omega) \\ E_y^g(\omega-\Omega) \end{bmatrix} \right.$$
$$\left. + e^{-i\Omega\tau} \begin{bmatrix} -E_y^*(\Omega) & -E_x^*(\Omega) \\ -E_x^*(\Omega) & E_y^*(\Omega) \end{bmatrix} \begin{bmatrix} E_x^g(\omega+\Omega) \\ E_y^g(\omega+\Omega) \end{bmatrix} \right), \quad (2)$$

where $\boldsymbol{E}(\Omega)$ and $\boldsymbol{E}^g(\omega)$ are complex amplitude spectra of multi-THz and gate pulses, respectively, $b$ is a coefficient proportional to the nonlinear susceptibility, and $\tau$ is the time delay between the gate and multi-THz pulses. SFG and DFG have been assumed to be much smaller than the incident light.

The gate pulse $\boldsymbol{E}^g(\omega)$ just before the EO crystal is described by using PEM and QWP Jones matrices,

$$\begin{bmatrix} E_x^g(\omega;\varphi) \\ E_y^g(\omega;\varphi) \end{bmatrix} = \begin{bmatrix} 1 & 0 \\ 0 & e^{i\Delta(\omega)} \end{bmatrix} R\left(\frac{\pi}{4}\right) \begin{bmatrix} 1 & 0 \\ 0 & e^{i\delta_0(\omega)\sin\varphi} \end{bmatrix} R\left(-\frac{\pi}{4}\right) \begin{bmatrix} 1 \\ 0 \end{bmatrix}, \quad (3)$$

where $\varphi$ is the PEM phase, $\delta_0(\omega)$ is the modulation depth of PEM considering frequency dependence, $\Delta(\omega)$ is phase retardation of QWP, and $R(\pm\pi/4)$ is the rotation matrix for angle of $\pm\pi/4$. We assume $\Delta(\omega) = \pi/2$ for an ideal QWP and $\Delta(\omega) = (\pi/2) \times \omega/\omega_0$ for a zero-order QWP. The modulation depth of PEM $\delta_0(\omega)$ is also assumed to have similar frequency dependence for ideal and actual situations. The gate pulse $E_x^{g,\text{WP}}(\omega,\tau;\varphi)$ before the Wollaston prism in balance detection is described as

$$\begin{bmatrix} E_x^{g,\text{WP}}(\omega,\tau;\varphi) \\ E_y^{g,\text{WP}}(\omega,\tau;\varphi) \end{bmatrix} = R\left(\frac{\pi}{4}\right) \begin{bmatrix} 1 & 0 \\ 0 & e^{i\Delta(\omega)} \end{bmatrix} R\left(-\frac{\pi}{4}\right) \begin{bmatrix} E_x^{g,\text{out}}(\omega,\tau;\varphi) \\ E_y^{g,\text{out}}(\omega,\tau;\varphi) \end{bmatrix}. \quad (4)$$

Then, the balanced signal $S(\tau;\varphi)$ becomes

$$S(\tau;\varphi) = \int d\omega \left( \left|E_x^{g,\text{WP}}(\omega,\tau;\varphi)\right|^2 - \left|E_y^{g,\text{WP}}(\omega,\tau;\varphi)\right|^2 \right). \quad (5)$$

After the numerical calculation of $S(\tau;\varphi)$, the same fitting procedure as the experiment was applied for each delay time $\tau$. The gate spectrum $\boldsymbol{E}^g(\omega)$ was assumed to be a Gaussian function with transform-limited pulse duration of 13 fs. The multi-THz field was also set as Gaussian with a center frequency of 30 THz and an FWHM of 20 THz. The effects of the imperfections in the PEM or QWP and the chirp in the gate pulse were evaluated in the simulations.

Figure 4(b) shows the results for the ideal case, where the PEM and QWP are perfectly achromatic, and the gate pulse is transform-limited. The retrieved $E_x$ almost coincides with the input waveform, and $E_y$ and $a_0$ are zero. The residue of the fitting, defined as the sum of the squared fitting errors and plotted on the right axis, was also quite small in this ideal case. Figure 4(c) shows the result when the frequency dependence of the PEM retardation is considered. Although $a_0$ has a finite value compared with that in Fig. 4(b), $E_x$ and $E_y$ are not affected. The zero order QWP is shown in Fig. 4(d). $E_x$ does not largely change, whereas $E_y$ shows a larger artifact than that of Fig. 4(c). The residue of the fitting was much larger than that shown in Fig. 4(c), indicating that the original fitting function [Eq. (1)] is no longer valid. Figures 4(e) and

4(f) show the results for a chirped gate pulse with a duration of 24 fs; the other conditions were the same as those in Figs. 4(c) and 4(d), respectively. In both cases, $E_x$ is slightly deformed because of the poor time resolution. When the QWP was achromatic (Fig. 4(e)), $E_y$ was almost zero and the residue was small. In contrast, $E_y$ shows a large artifact, and the residue becomes large when the QWP is of the zero order (Fig. 4(f)). From these simulations, we confirmed that the POMEOS fitting function, including the offset term $a_0$ is also valid for a 10-fs-class broadband gate pulse as long as achromatic QWP is used. We consider that our experimental results correspond to a mixture of slight chirp in the gate pulse and small imperfections in the achromatic QWP (~0.004$\lambda$).

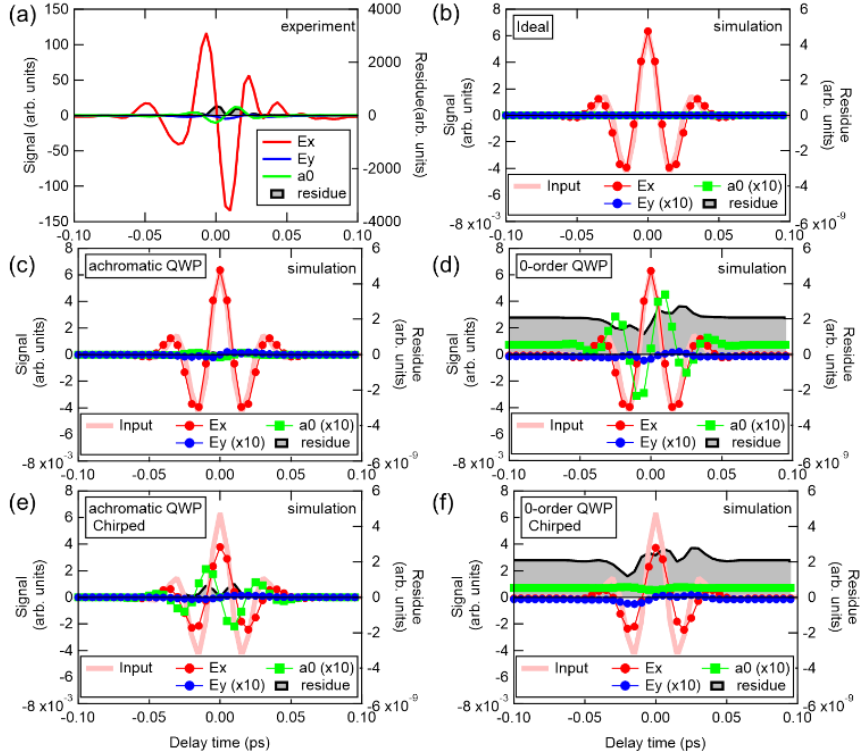

Fig. 4. (a) An experimental result for a linearly polarized pulse. Red: $E_x$, blue: $E_y$, green: offset term, black shaded: residue in fitting (right axis). (b–f) Results of numerical simulations. (b) Ideal case with achromatic PEM and QWP, (c) frequency-dependent PEM and achromatic QWP, and (d) frequency-dependent PEM and QWP. (e) and (f) are calculated for a chirped gate pulse, with the other parameters the same as (c) and (d), respectively.

## 4. Conclusions

We demonstrated the characterization of time-domain waveform of a multi-THz pulse as vector quantities. The POMEOS technique was successfully extended to the multi-THz region with a broadband gate pulse shorter than 15 fs. The fitting function in POMEOS was modified with an offset term and its validity was confirmed through numerical simulations. Owing to the fast polarization modulation, we demonstrated high accuracy and precision of approximately 1 mrad with only 3 s of data accumulation. Our method evaluates a multi-THz electric field vector, which can be applied to ultrafast time-resolved measurements of the chirality of molecules, metamaterials, or off-diagonal response functions in magnetic systems away from equilibrium. Our method can also be important in Floquet engineering using a designed complex vector field, such as counter-rotating bicircular polarized light [14–16] by utilizing the vector field-shaping technique [36].

**Funding.** JST PRESTO (JPMJPR2006), and the Ministry of Education, Culture, Sports, Science and Technology, Quantum Leap Flagship Program (JPMXS0118068681).

**Disclosures.** The authors declare no conflicts of interest.

**Data availability.** The data for this study are available from the corresponding authors upon request.